\documentclass[article,preprint]{elsarticle}
\usepackage{a4wide}
\usepackage{natbib} 			
\usepackage{graphicx}
\usepackage{amsfonts}
\usepackage{amsmath}
\usepackage{booktabs}
\usepackage{epsfig}
\usepackage{hyperref}
\hypersetup{citecolor=black,linkcolor= black}

\begin{document}
\title{On the interplay between short and long term memory in the power-law cross-correlations setting}
\author{Ladislav Kristoufek}
\ead{kristouf@utia.cas.cz}
\address{Institute of Information Theory and Automation, Academy of Sciences of the Czech Republic, Pod Vodarenskou Vezi 4, 182 08, Prague 8, Czech Republic\\
Institute of Economic Studies, Faculty of Social Sciences, Charles University, Opletalova 26, 110 00, Prague 1, Czech Republic
}

\begin{abstract}
We focus on emergence of the power-law cross-correlations from processes with both short and long term memory properties. In the case of correlated error-terms, the power-law decay of the cross-correlation function comes automatically with the characteristics of separate processes. Bivariate Hurst exponent is then equal to an average of separate Hurst exponents of the analyzed processes. Strength of short term memory has no effect on these asymptotic properties. Implications of these findings for the power-law cross-correlations concept are further discussed.
\end{abstract}


\journal{Physica A}

\maketitle


\section{Introduction}

Existence of statistically significant power-law cross-correlations between various series is a fascinating phenomenon important for modeling and forecasting time series. Several processes that possess such long-term correlations have been proposed in the literature. The most frequently discussed and applied ones are multivariate generalizations of the well-established fractionally integrated ARMA processes (usually labeled as FARIMA and ARFIMA) -- VARFIMA or MVARFIMA processes -- and fractional Gaussian noise processes or fractional Brownian motions, which are their integrated version (these are labeled as fGn and fBm in the literature, respectively) \cite{Lobato1997,Ravishanker1997,Martin1999,Nielsen2004,Shimotsu2007,Sela2009,Tsay2010,Nielsen2011}. Construction of the multivariate ARFIMA process implies that the bivariate Hurst exponent is the average of the separate Hurst exponents \citep{Nielsen2011} and the same property holds for the fractional Brownian motion \citep{Amblard2011}. The long-range cross-correlations thus simply arise from the specification of these processes. However, most of the studies focus on a specific case when both studied series are themselves power-law correlated leaving aside a possibility that one of the processes is indeed not power-law correlated. Here, we focus on two specific cases -- a pair of power-law correlated processes, and a combination of a power-law correlated and an exponentially correlated processes -- and compare their properties in the power-law cross-correlations framework. For a sake of simplicity and straightforward results, we stick to the ARFIMA setting usually followed in the multidisciplinary physics literature \cite{Podobnik2008,Podobnik2008a,Zhou2008,Jiang2011,Kristoufek2011,Kristoufek2013}.

\section{Power-law cross-correlated processes}

Power-law cross-correlated processes are usually defined via a power-law decay of a cross-correlation function. Specifically, if the cross-correlation function $\rho_{xy}(k)$ between processes $\{x_t\}$ and $\{y_t\}$ decays as $\rho_{xy}(k) \propto k^{2H_{xy}-2} \equiv k^{-\gamma_{xy}}$ with lag $k \rightarrow +\infty$, we say that the processes are power-law cross-correlated. The characteristic bivariate Hurst exponent $H_{xy}$ has a similar interpretation as the univariate one so that if $H_{xy}>0.5$, the processes are cross-persistent, and if $H_{xy}<0.5$, the processes are cross-antipersistent. Both $H_{xy}$ and $\gamma_{xy}$ (in addition to other parameters) are used in the literature interchangeably, depending on an initial setting of the correlation structure.

\subsection{Correlated ARFIMA processes}

We start with ARFIMA processes with correlated error-terms simply structured as two ARFIMA(0,$d$,0) processes with parameters $d_1$, $d_2$, $a_n(d)=\frac{\Gamma(n+d)}{\Gamma(n+1)\Gamma(d)}$ and a specific correlation structure:
\begin{gather}
\label{eq1}
x_t=\sum_{n=0}^{\infty}{a_n(d_1)\varepsilon_{t-n}} \\
y_t=\sum_{n=0}^{\infty}{a_n(d_2)\nu_{t-n}} \nonumber \\
\langle \varepsilon_t \rangle = \langle \nu_t \rangle = 0 \nonumber\\
\langle \varepsilon_t^2 \rangle = \sigma_{\varepsilon}^2 < +\infty \nonumber\\
\langle \nu_t^2 \rangle = \sigma_{\nu}^2 <+\infty \nonumber\\
\langle \varepsilon_t\varepsilon_{t-n} \rangle = \langle \nu_t\nu_{t-n} \rangle = \langle \varepsilon_t\nu_{t-n} \rangle = 0\text{ for }n \ne 0 \nonumber\\
\langle \varepsilon_t\nu_t \rangle = \sigma_{\varepsilon\nu} <+\infty \nonumber.
\label{eq:ARFIMA_varcovar}
\end{gather}
Note that both processes are stationary \cite{Sowell1992,Bertelli2002,Samorodnitsky2006}. Cross-power spectrum $f_{xy}(\lambda)$ with frequency $0<\lambda \le \pi$ of the two processes can be written as
\begin{multline}
\label{eq:spectrum_ARFIMA_text}
f_{xy}(\lambda)=\frac{\sigma_{\varepsilon\nu}}{2\pi}\sum_{k=0}^{\infty}\sum_{l=0}^{\infty}{a_k(d_1)a_l(d_2)\exp(i(k-l)\lambda)}=
\frac{\sigma_{\varepsilon\nu}}{2\pi}\left(1-\exp(i\lambda)\right)^{-d_1}\left(1-\exp(-i\lambda)\right)^{-d_2}. \nonumber
\end{multline}
To show whether the processes are power-law cross-correlated, we need to inspect an asymptotic behavior of the cross-correlation function $\rho_{xy}(n)$. Using the inverse Fourier transform of the cross-power spectrum, we can write the $n$th cross-correlation as
\begin{equation}
\rho_{xy}(n)=\frac{\sigma_{\varepsilon\nu}}{2\pi\sigma_x\sigma_y}\sum_{k=0}^{\infty}\sum_{l=0}^{\infty}{a_k(d_1)a_l(d_2)\int_{-\pi}^{\pi}\exp(i(n+k-l)\lambda)d\lambda}.
\label{eq:rhon}
\end{equation}
Now, using the definition and properties of the Dirac delta function \citep{Dirac1958}, we can rewrite the cross-correlation function in Eq. \ref{eq:rhon} as
\begin{equation}
\label{eqDirac}
\rho_{xy}(n)=\frac{\sigma_{\varepsilon\nu}}{\sigma_x\sigma_y}\sum_{k=0}^{\infty}\sum_{l=0}^{\infty}{a_k(d_1)a_l(d_2)\delta(n+k-l)}=\frac{\sigma_{\varepsilon\nu}}{\sigma_x\sigma_y}\sum_{k=0}^{\infty}{a_k(d_1)a_{n+k}(d_2)} \nonumber
\end{equation}
and follow with
\begin{equation}
\sum_{l=0}^{\infty}{a_k(d_1)a_l(d_2)\delta(n+k-l)}=\sum_{l=0}^{\infty}{a_k(d_1)a_l(d_2)\delta(l-n-k)}=\sum_{l=0}^{\infty}{a_k(d_1)a_{n+k}(d_2)} \nonumber
\end{equation}
where $t=l$ and $a=n+k$. We can now rewrite $a_j(d)$ with a use of the Beta function so that 
\begin{equation}
a_j(d)=\frac{\Gamma(j+d)}{\Gamma(j+1)\Gamma(d)}=\frac{1}{kB(k,d)}. \nonumber
\label{eq:aj}
\end{equation}
Using Stirling's approximation of the Beta function $B(\bullet,\bullet)$ for fixed $d$ and $j \rightarrow +\infty$, we get
\begin{equation}
\label{eq:Stirling1}
a_j(d) \approx \frac{1}{j}\frac{1}{\Gamma(d)j^{-d}}=\frac{j^{d-1}}{\Gamma(d)}.
\end{equation}
Since we are interested in the asymptotic behavior, we can use Eq. \ref{eq:Stirling1} and follow with
\begin{equation}
\rho_{xy}(n) \approx \frac{\sigma_{\varepsilon\nu}}{\sigma_x\sigma_y\Gamma(d_1)\Gamma(d_2)}\sum_{k=0}^{\infty}{k^{d_1-1}(n+k)^{d_2-1}} \nonumber
\end{equation}
given that $d_1,d_2,k,n+k>0$. Approximating the infinite sum with a definite integral, we can write
\begin{multline}
\label{eq:ARFIMA1}
\rho_{xy}(n) \approx \frac{\sigma_{\varepsilon\nu}}{\sigma_x\sigma_y\Gamma(d_1)\Gamma(d_2)}\int_{0}^{\infty}{k^{d_1-1}(n+k)^{d_2-1}dk}=\\
=\frac{\sigma_{\varepsilon\nu}}{\sigma_x\sigma_y\Gamma(d_1)\Gamma(d_2)}n^{d_1+d_2-1}\frac{\Gamma(d_1)\Gamma(1-d_1-d_2)}{\Gamma(1-d_2)}=\\
= \frac{\sigma_{\varepsilon\nu}\Gamma(1-d_1-d_2)}{\sigma_x\sigma_y\Gamma(1-d_2)\Gamma(d_2)}n^{d_1+d_2-1} \propto n^{d_1+d_2-1}=n^{-(1-d_1-d_2)} \nonumber
\end{multline}
given that $d_1+d_2<1$ and $n>0$. Therefore, given that $\sigma_{\varepsilon\nu}\ne 0$, the power-law cross-correlations emerge regardless of the level of correlation between error-terms $\{\varepsilon_t\}$ and $\{\nu_t\}$ as long as it is non-zero. Using the relationship between fractional differencing parameter and Hurst exponent $d=H-0.5$, we have
\begin{equation}
H_{xy}=1-\frac{\gamma_{xy}}{2}=1-\frac{1-d_1-d_2}{2}=1-\frac{-(H_x+H_y)+2}{2}=\frac{H_x+H_y}{2}.
\label{eq:ARFIMA_H}
\end{equation}
The bivariate Hurst exponent $H_{xy}$ is thus an average of the separate Hurst exponents $H_x$ and $H_y$ regardless the correlation between error-terms as long as it remains non-zero. This also covers the case showed in Ref. \cite{Podobnik2009} for two ARFIMA processes with the identical error-terms. We now turn to the combination of short and long term memory.

\subsection{Combination of AR and ARFIMA processes}

In the univariate case, distinguishing between short and long term memory is evident from the properties of the auto-correlation function. To see how these two types of memories interact in the bivariate setting, we investigate the case when one of the processes is long-range dependent, the other is short-range dependent and their error-terms are pairwise correlated. Let's have ARFIMA process $\{x_t\}$ and AR(1) process $\{y_t\}$ defined as
\begin{gather}
\label{eq2}
x_t=\sum_{n=0}^{\infty}{a_n(d_1)\varepsilon_{t-n}} \\
y_t=\theta y_{t-1}+\nu_t \nonumber
\label{eq:ARFIMAAR}
\end{gather}
with $|\theta|<1$. Moments of the error-terms are specified as for the previous case and the processes are thus stationary \cite{Wei2006}. The cross-power spectrum has the following form
\begin{multline}
f_{xy}(\lambda)=\frac{\sigma_{\varepsilon\nu}}{2\pi}\sum_{k=0}^{\infty}\sum_{l=0}^{\infty}{a_l(d_1)\theta^k\exp(i(k-l)\lambda)}=
\frac{\sigma_{\varepsilon\nu}}{2\pi}\left(1-\exp(-i\lambda)\right)^{-d_1}\left(1-\theta\exp(i\lambda)\right)^{-1}. \nonumber
\label{eq:spectrum_ARFIMA_AR1_text}
\end{multline}
Using the inverse Fourier transform, we get
\begin{equation}
\rho_{xy}(n)=\frac{\sigma_{\varepsilon\nu}}{2\pi\sigma_x\sigma_y}\sum_{k=0}^{\infty}\sum_{l=0}^{\infty}{a_l(d_1)\theta^k\int_{-\pi}^{\pi}\exp(i(n+k-l)\lambda)d\lambda}. \nonumber
\end{equation}
Again, we use the definition of Dirac's delta function and its properties to get
\begin{equation}
\rho_{xy}(n)=\frac{\sigma_{\varepsilon\nu}}{\sigma_x\sigma_y}\sum_{k=0}^{\infty}\sum_{l=0}^{\infty}{a_l(d_1)\theta^k\delta(n+k-l)}=\frac{\sigma_{\varepsilon\nu}}{\sigma_x\sigma_y}\sum_{k=0}^{\infty}{a_{n+k}(d_1)\theta^k}. \nonumber
\end{equation}
Using the Stirling's approximation and approximating the infinite sum by the definite integral, we get
\begin{equation}
\rho_{xy}(n)\propto \int_{0}^{\infty}{(n+k)^{d_1-1}\theta^kdk}=\theta^{-n}\Gamma(d_1,-n\log\theta)(-\log\theta)^{-d_1} \nonumber
\end{equation}
where $\Gamma(\bullet,\bullet)$ is the incomplete upper Gamma function \citep{Wall1948}. Using the approximation of the incomplete upper Gamma function \cite{Blahak2010}, we can write
\begin{multline}
\rho_{xy}(n)\propto \theta^{-n}(-\log\theta)^{-d_1}(-n\log\theta)^{d_1-1}\exp(n\log\theta)\\
=\theta^{-n}\theta^{n}(-\log\theta)^{-d_1}(-\log\theta)^{d_1+1}n^{d_1-1}\propto n^{d_1-1} \nonumber
\end{multline}
Therefore, we have
\begin{equation}
H_{xy}=1-\frac{\gamma_{xy}}{2}=1-\frac{1-d_1}{2}=1-\frac{-H_x+1.5}{2}=\frac{H_x+0.5}{2}
\end{equation}
which is perfectly in hand with Eq. \ref{eq:ARFIMA_H} for $H_y=0.5$, i.e. the process $\{y_t\}$ is not long-range dependent with $d_2=0$. Note that the asymptotic relationship is again independent of $\sigma_{\varepsilon\nu}$ as long as $\sigma_{\varepsilon\nu}\ne0$.

\section{Discussion and conclusions}

Two types of processes generating power-law cross-correlations have been studied in detail. Importantly, such cross-correlations very easily arise from a very simple specification of the separate processes. As long as the error-terms are correlated, the power-law decay of the cross-correlation function emerges from the correlation structure of the separate processes. Moreover, we have presented that even if one of the analyzed processes is only short-range (exponentially) correlated, the long term memory of the other processes dominates and the processes together form a power-law cross-correlated pair. This is true regardless of a strength of the short term memory component. 

These theoretical results are extremely important for empirical studies of power-law cross-correlations across various disciplines. They imply that the usually reported result of $H_{xy}>0.5$ connected with $H_{xy}\approx \frac{1}{2}(H_x+H_y)$ is not necessarily a sign of complex dependence between the analyzed series but it might simply emerge from the fact that at least one of the series is power-law correlated and the error-terms of the processes are at least somehow correlated. As the former case -- power-law correlated separate processes -- is quite frequent, we focus on the latter one -- correlated error-terms. This brings us to a very understanding and interpretation of the error-term in statistical analysis. Even though there are many approaches, we stick to the two most common ones. Error-term can be understood primarily as a measurement error which brings uncertainty into a well defined model. Possible correlation between measurement errors of two series is thus usually not expected. Note that such definition is prevalent in experimental studies. However, error-terms in time series analysis are more frequently taken as unexpected innovations to the system or sometimes also as an unpredictable information flow. This is nicely illustrated in definitions of ARFIMA and AR processes in Eqs. \ref{eq1} and \ref{eq2}. The unpredictable innovations flow into the system and they get translated into final processes $\{x_t\}$ and $\{y_t\}$ based on their memory characteristics. Such innovations can either have a long-term or a short-term effect on the overall dynamics of a given process which is characterized by a specification as either AR or ARFIMA process (or generally many other possible specifications).

For the time series analysis, the latter understanding of error-terms prevails. In many systems, it is meaningful to expect correlated error-terms. As the power-law correlations and cross-correlations are heavily examined in economic and financial series, let us illustrate the concept on a simple example from econophysics. There, examination of cross-correlations between returns, volatility (riskiness of an asset) and traded volume is very popular. Now assume that an unexpected (not necessarily extreme) negative event occurs, e.g. during a quarterly profit announcement. Such event is negative information coming into the system and it is thus an innovation, an error-term (or a part of it) for the examined processes. The information in turn affects all three studied variables -- price and thus also returns react negatively, volatility increases due to a magnified uncertainty and traders become more active as they try to rebalance their positions according to a new market situation. Therefore, all three variables react to the same impulse and their error-terms are thus correlated. Moreover, volatility and traded volume are power-law correlated which means that emerging power-law cross-correlations between returns, volatility and traded volume are present automatically as shown in this paper.

However, this should not discard the power-law cross-correlations analysis as a concept. Our findings stress that the analysis needs to be complete, without only partial results of the bivariate Hurst exponent. Empirical analyses should not be reported in a purely technical manner but they should be put into a correct context of the examined series. Another direction of research also further opens for processes with $H_{xy} \ne \frac{1}{2}(H_x+H_y)$ which have been studied only marginally \cite{Podobnik2008a,Kristoufek2013,Sela2012}. Branch of the power-law cross-correlations thus still remains an open field with numerous possibilities for future research.

\section*{Acknowledgements}

The author would like to thank the anonymous referees for valuable comments and suggestions which helped to improve the paper significantly. Support from the Czech Science Foundation under project No. 14-11402P and Grant Agency of the Charles University in Prague under project No. 1110213 is gratefully acknowledged.


\section*{References}
\bibliography{Bibliography}
\bibliographystyle{unsrt}

\end{document}